\documentclass[aps,pre,twocolumn,floatfix,superscriptaddress,10pt]{revtex4-2}
\usepackage{amsfonts,amssymb,amsmath,amsthm}
\usepackage{graphicx}
\usepackage{color}
\usepackage{hyperref}
\newcommand{\Reff}{R_{\textrm{eff}}}

\begin{document}
\title{Electrical Transport in Tunably-Disordered Metamaterials}

\author{Caitlyn Obrero}
\affiliation{Department of Materials Science and Engineering, North Carolina State University, Raleigh, NC, USA}

\author{Mastawal Tirfe}
\affiliation{Department of Physics, North Carolina State University, Raleigh, NC, USA}

\author{Carmen Lee}
\affiliation{Department of Physics, North Carolina State University, Raleigh, NC, USA}

\author{Sourabh Saptarshi}
\affiliation{Department of Mechanical and Aerospace Engineering, North Carolina State University, Raleigh, NC, USA} 

\author{Christopher Rock}
\affiliation{Center for Additive Manufacturing and Logistics, North Carolina State University, Raleigh, NC, USA}

\author{Karen E. Daniels}
\affiliation{Department of Physics, North Carolina State University, Raleigh, NC, USA}

\author{Katherine A. Newhall}
\affiliation{Department of Mathematics, University of North Carolina at Chapel Hill, Chapel Hill, NC, USA}

\date{\today} 

\begin{abstract}
Naturally occurring materials are often disordered, with their bulk properties being challenging to predict from the structure,  due to the lack of underlying crystalline axes.  
In this paper, we develop a digital pipeline from algorithmically-created configurations with tunable disorder to 3D printed materials, as a tool to aid in the study of such materials, using electrical resistance as a test case.
The designed material begins with a random point cloud that is iteratively evolved using Lloyd's algorithm to approach uniformity, with the points being connected via a Delaunay triangulation to form a disordered network metamaterial. 
Utilizing laser powder bed fusion additive manufacturing with stainless steel 17-4 PH and titanium alloy Ti-6Al-4V, we are able to experimentally measure the bulk electrical resistivity of the disordered network. 
The effective resistance of the structure calculated from the 
combinatorial weighted graph Laplacian 
is in good agreement with experimental data.
However, the effective resistance is sensitive to anisotropy and global network topology, preventing a single network statistic or disorder characterization from predicting global resistivity.

\end{abstract}


\maketitle

\section{Introduction}

Research on metamaterials, whether mechanical or photonic, has focused primarily on ordered structures \cite{mueller_energy_2019, turpin_reconfigurable_2014, bauer_nanolattices_2017, askari_additive_2020, surjadi_mechanical_2019, ma_deep_2021}, where the bulk properties are inherited from both the constituent material and the connectivity of the lattice. In contrast, the ubiquity of disorder within naturally-occurring materials raises the question of how bulk properties arise when the underlying structure is no longer composed along well-defined crystalline axes.
A particularly interesting case is that of hyperuniform structures, 
for which large-scale density fluctuations are anomalously suppressed compared to those in typical disordered systems \cite{torquato_hyperuniform_2018,chieco_characterizing_2017}.
Such materials 
are predicted to thereby have special transport properties such as a higher critical current in type-II superconductors, possibly due to fewer weak links, and changes in the exponent describing diffusion processes in disordered media \cite{torquato_hyperuniform_2018}. The potential use of off-lattice, disordered metamaterials --- as has recently attracted interest \cite{siedentop_stealthy_2024, zaiser_disordered_2023,  sniechowski_heterogeneous_2015} --- has a key advantage of providing a much larger parameter space to explore than lattices alone can provide. 

\begin{figure*}
\includegraphics[width=0.95\linewidth]{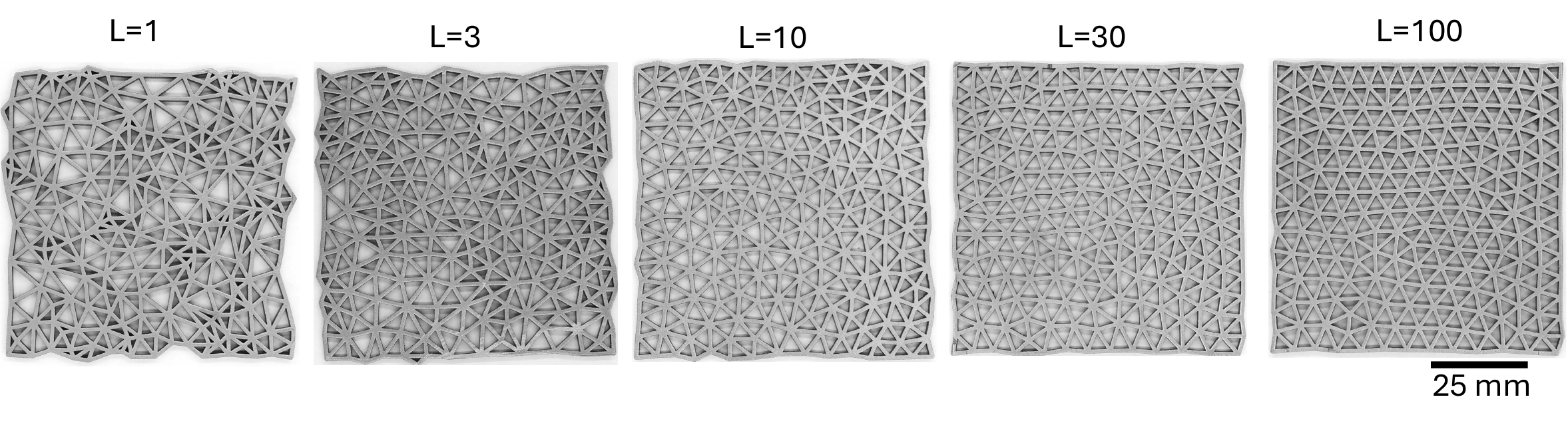}
\caption{Photos of additively-manufactured Ti-6Al-4V samples created from  $N=200$ nodes, subjected to $L=1,3,10,30,100$ iterations of Lloyd's algorithm. A movie of the evolving point cloud from $L=1$ to $L=100$ is available at \cite{youtube}.}
\label{fig:4lloyds}
\end{figure*}

In order to explore the effects of disorder on transport properties, we consider the test case of electrical resistance within a tunably-disordered network of metal beams (see Fig.~\ref{fig:4lloyds}) created from Lloyd's algorithm \cite{lloyd_least_1982}, which is known to generate hyperuniform point clouds in the many-iteration limit \cite{klatt_universal_2019,hong_dynamical_2021}. 
The choice of electrical resistance provides both a quantity easily measured in the lab, and an exact prediction via the combinatorial graph Laplacian \cite{newman_networks_2018}. When writing the configuration of beams as a network, the edge weights encode the electrical resistance of each beam, which is proportional to its length. Using the graph Laplacian to find an effective resistance, $\Reff$, across the entire network is equivalent to solving the full equations (Kirchhoff's laws), which we find is quite sensitive to small configurational changes in the network.

Our numerical tests and physical samples (see Fig.~\ref{fig:4lloyds}) are both created via the same procedure, beginning from a disordered point cloud that is evolved under Lloyd's algorithm to a more-ordered (approaching hyperuniform \cite{klatt_universal_2019,hong_dynamical_2021}) 
configuration. This allows us to use numerical methods to explore the variability between configurations created by different realizations of the same process, as well as across systems of different sizes. For select configurations, we use laser powder bed fusion additive manufacturing (L-PBF AM) to create physical metal alloy samples \cite{sames_metallurgy_2016,debroy_additive_2018}, providing a validation of the approach.

For each configuration, we characterize the disorder via the edge-length distribution, degree distribution, and entropy (degree or edge). 
We find that the graph Laplacian accurately reproduces the measured effective (bulk) resistance $\Reff$ for each printed network with no free parameters, only the known resistivity of the printed material. In addition, the observed variability in $\Reff$ within an ensemble of configurations produced under the same protocol is largest for the most disordered systems, which are also those with the largest variance in edge length. In all cases, we observe that the predicted current on specific edges depends both on network topology and the direction of applied voltage, with significant anisotropy present.

\section{Methods}

Our methods include designing, fabricating, and experimentally measuring tunably-disordered network materials.
In Sec.~\ref{sec:network_design}, we describe our process of generating a sequence of network configurations that define the material we are making.  This combines Lloyd's algorithm to iteratively move the network nodes with a Delaunay triangulation to connect the nodes with edges.
In Sec.~\ref{sec:AM}, we describe how these configurations are manufactured using laser powder bed fusion.    
In Sec.~\ref{sec:measure}, we explain how effective resistance is measured experimentally, which will be compared to the theoretically computed effective resistance based on Kirchhoff's Laws in Sec.~\ref{sec:Reff}.

\subsection{Generating configurations\label{sec:network_design}}

Our tunably-disordered network materials were constructed from point clouds (nodes) that were connected together by edges using a  Delaunay triangulation subjected to multiple iterations of Lloyd's algorithm \cite{lloyd_least_1982} to make them progressively more ordered.  Five examples, from $L=1$ to $L=100$ iterations, are shown in Fig.~\ref{fig:4lloyds}, each created using the procedure described below.

Each series of related configurations was started by selecting $N$ points drawn from a uniform distribution within a unitless $2000\times 2000$ bounding box. This collection of $N$ points is referred to as a point cloud.
We first generated a Voronoi tessellation within the pre-defined bounding box.  Each Voronoi polygon contains all the points in the bounding box that are closer to one particular point cloud point than to any other.
We then performed one iteration of Lloyd's algorithm, updating the location of the point cloud points to the centroid of each Voronoi polygon, following the methods of \cite{Swarm_Lab}.
This first iteration was denoted $L=1$. We then recomputed the new Voronoi tessellation in the bounding box and performed Lloyd's algorithm again, up to $L=100$ iterations. 
Our disordered lattices were constructed from the Delaunay triangulation of the point cloud at any step of this process. The Delaunay triangulation connects any two point cloud points if their respective Voronoi polygons share an edge. 
The code for generating these configurations is available on GitHub \cite{GitHub_config_generation}.

 Starting from one initial point cloud chosen uniformly, subsequent iterations 
of Lloyd's algorithm approaches crystallinity \cite{klatt_universal_2019, hain_low-temperature_2020, hong_dynamical_2021}. As we will see in Sec.~\ref{sec:netchar}, this change in uniformity happens quickly at first, then progresses approximately logarithmically. In our numerical analyses, we use 20 different initial point clouds at each of $N=100, 200, 300,$ and $500$, analyzed from $L=0$ (the original point cloud) to 100. In our experiments, we selected one initial point cloud with $N=200$ points, and printed physical samples at iteration counts $L=1$, 3, 10, 30, and 100.

\subsection{Sample fabrication \label{sec:AM}}

As shown in Fig.~\ref{fig:4lloyds}, we selectively printed one realization of the $N=200$ network configurations using laser powder bed fusion in two materials: stainless steel 17-4 PH and titanium Ti-6Al-4V. Each of the five Delaunay triangulation connected configurations ($L=1, 3, 10, 30, 100$) was converted into an STL file for 3D printing using computer-aided design (CAD) software and generative modeling. All connecting beams were designed with rectangular cross sections 1~mm wide and 3~mm tall and are printed within a bounding box that has dimensions 75~mm $\times$ 75~mm ($\pm 2$~mm).  The sample thickness for printing was set larger than 3~mm within the Materialize Magics slicing software by extruding the STL geometry in the vertical direction to the desired height; samples are then sliced to a 3~mm thickness as detailed below.

The fabrication relies on power bed fusion machine parameters described by the volumetric energy density, $E_V$, a common method of representing variables associated with input energy and related process parameters.  The volumetric energy density $E_V$  is given by 
\begin{equation}\label{eq:energydensity}
    E_{V} = \frac{P}{vht}
\end{equation}
with laser power $P$, 
laser beam speed $v$, 
laser hatch spacing $h$, 
and powder layer thickness $t$ \cite{simchi_direct_2006}.  

Three different printing processes were used for the samples, to test two printers on the two materials. (1) The titanium samples were printed using  Ti-6Al-4V powder on a modified GE Concept Laser Mlab 100R L-PBF machine, with a 100~W Nd:YAG fiber laser with 1070 nm wavelength and 50~$\mu$m spot size. The following melt parameters were selected after mapping the process space to optimize material density: power $P= 90$~W, beam speed  $v = 825$~mm/s, raster hatch spacing $h = 0.08$~mm, and layer thickness  $t = 0.03$~mm. This gives an energy density $E_V = 45 \, \text{J/mm}^3$. (2) Most stainless steel samples ($L=1, 10, 30, 100$) were printed using the EOS M290 with a 400~W Yb-fiber laser with a 1060~nm wavelength and an 80 $\mu$m spot size. The stainless steel networks on EOS M290 are fabricated using a layer thickness $t = 0.04$~mm, laser power $P= 220$~W, beam speed $v =755$~mm/s, and raster hatch spacing $h = 0.10$~mm with an energy density of $73 \, \text{J/mm}^3$ \cite{shrestha_effects_2018}, with one exception.  (3) The steel $L=3$ sample was printed on the Concept Laser MLab with $E_V = 66 \, \text{J/mm}^3$, to account for laser differences between the two machines. Both steel and titanium alloy powders were nominally $15-45 \, \mu$m in size with spherical morphology. 

The network geometries were fabricated to be $4-10$ mm tall with the build direction normal to the 2D network. The Ti-6Al-4V specimens were sectioned from the plate in the as-built condition. Because the 17-4 PH specimens and build plate experienced significant plate warping, post-fabrication the parts were heated to 1040{\textdegree}C in an Ar environment. This is referred to as a solution heat treatment condition, which relieves the stress observed after fabrication. The samples were sectioned using a Mitsubishi FA10S EDM (electrical discharge machining). During method development, two titanium configurations were harvested by leveling the EDM to the top surface of the printed network, and then making a slice to section the geometry from the build plate, leaving the top surface intact.  All subsequent specimens were leveled by first sectioning $0.5-1.0$~mm below the top surface, followed by harvesting the network.   The samples were cut to $(3.0 \pm 0.2)$~mm for steel and $(3.0 \pm 0.4)$~mm for titanium. Depending on the build plate size, post-fabrication build plate condition, and total print height, a total of 1 to 2 samples of each $L$ value per build was extracted. We cleaned each sample after sectioning using diluted Simple Green and Citranox solutions, respectively, followed by bead blasting to remove any lubricant, EDM recast, and residual oxide layer from the cutting operation or heat treatment. Small representative network sections  of each material were sectioned and metallurgically prepared using 600 grit SiC until flat, then polished using 3 micron diamond paste. In optical microscopy of sample sections, we observed dense network beams with low  porosity.

\subsection{Electrical measurements \label{sec:measure}}

\begin{figure}
\includegraphics[width=\linewidth]{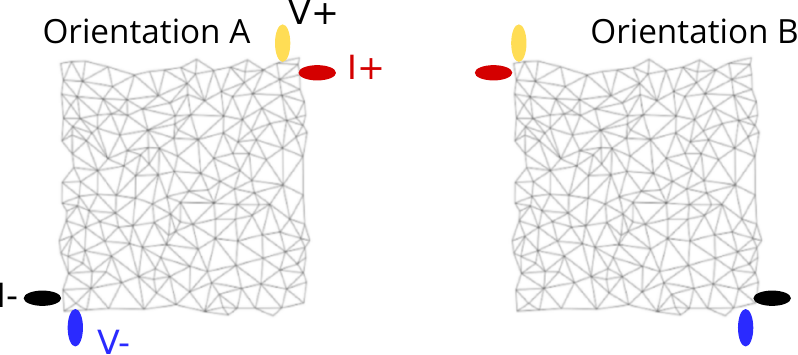}
\caption{Schematic of the two measurement orientations (named Orientation A and B) for the force current $I$ and measured voltage $V$ in the 4-point probe resistance measurements, with the orientation of the clamped nodes being either SW-NE or SE-NW, respectively. }
\label{fig:configuration}
\end{figure}

We conducted 4-point probe electrical resistance measurements using a Keithley 2450 SourceMeter on a total of 4 sets of 4 steel samples and 1 set of 5 titanium samples. In each case, we start from a single point cloud containing $N=200$ points, and iterate Lloyd's algorithm $L = 1, 3, 10, 30, 100$ times to achieve either 4 (omitting $L=3$) or 5 samples to test. In the results that follow, we report on all but the $L=3$ steel samples, for which the printing parameters were sufficiently different that a direct comparison is not appropriate. 
For each sample, two measurement orientations (A and B) were used, in order to probe whether our results depend on the direction the current is applied (see Fig.~\ref{fig:configuration}). The instrument applied a force current $I=100$~mA  on two leads, and measured voltage $V$ on two independently-connected leads for a duration of 500~ms. Each sample and orientation combination was repeated 6 times, making fresh connections of the 4 probes to the clamped corners for each repeated measurement. Resistance measurements for each sample are reported in Ohms ($\Omega$), corresponding to a bulk measurement.

To measure the material resistivity for both materials (17-4PH steel and Ti-6Al-4V), we printed serpentine specimens using the same L-PBF process and post-processing described above. Three specimens of each material were cut to $\sim4.5$~mm tall with a 1~mm$\times$1~mm cross-section and 
lengths 27.28, 30.28, and 60.64~mm. Following the ASTM B193 standard protocol for measuring electrical resistivity of metallic electrical conductor materials, we performed 4-point probe measurements; as described above for the network samples, we placed the  probes at the ends of the specimens and reconnected them for repeated measurements. The average resistivity for 17-4PH steel and Ti-6Al-4V were 102.68~$\mu \Omega\cdot$~cm and 195.40~$\mu \Omega\cdot$~cm, respectively.

\section{Results}

\subsection{Network characterizations \label{sec:netchar}}

Each configuration can be written as a network, with the nodes representing the center of each Voronoi cell, connected by edges (the Delaunay triangulation) weighted by either the length of the beams or its reciprocal. Weighting by the lengths is equivalent to weighting by the electrical resistance of the beams, and the reciprocal represents the electrical conductance. Both weightings are made under the assumption of dense, uniformly-printed materials.

\begin{figure}
\includegraphics[width=0.75\linewidth]{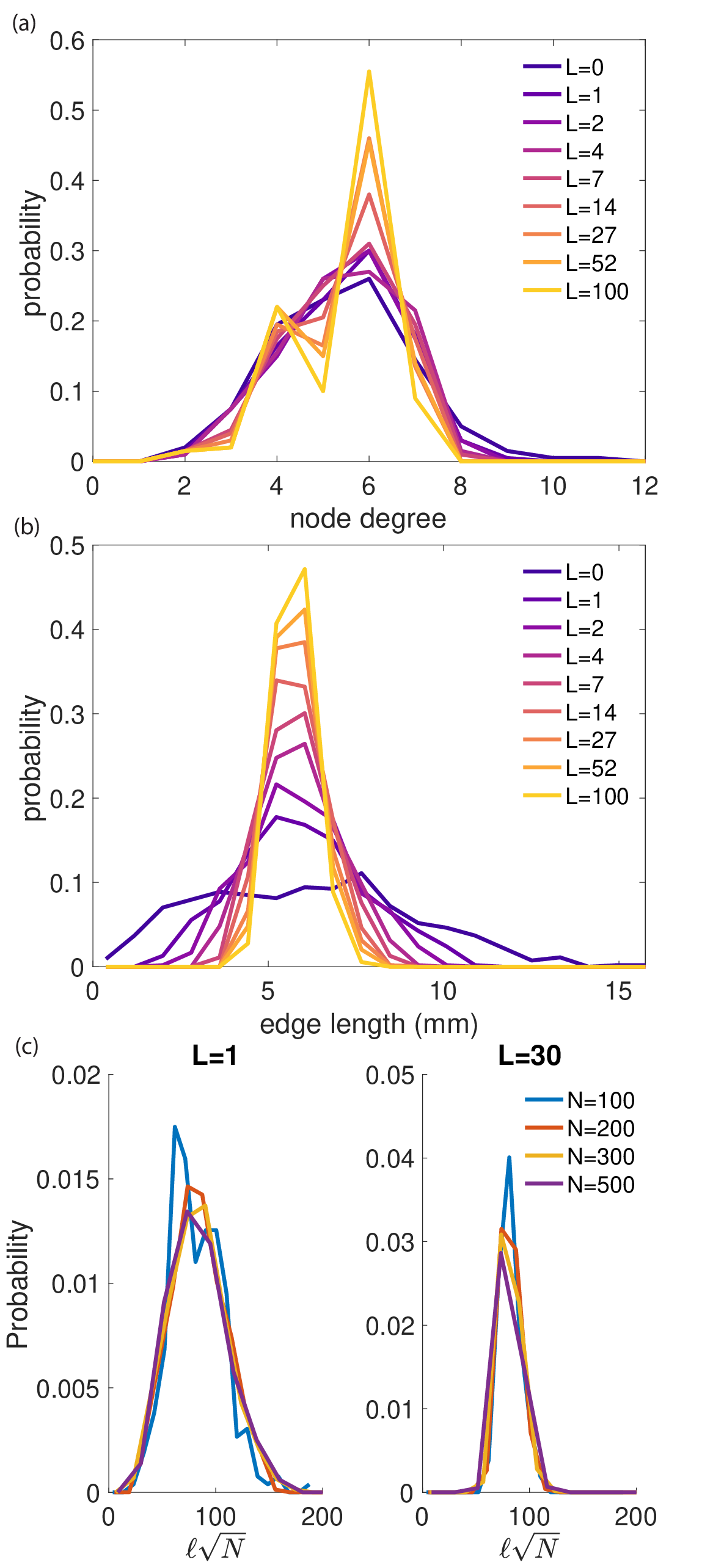}
\caption{For a single realization of an $N=200$ network subjected to Lloyd's iterations $L$, its (a) degree distribution $p_k$ and (b)  probability distribution function $f(\ell)$. (c) Edge-length distributions scaled by $\sqrt{N}$ at Lloyd's iterations $L=1$ and $L=30$ for networks with increasing number of nodes $N$. }
\label{fig:disorder}
\end{figure}

As seen in Fig.~\ref{fig:4lloyds}, the networks  become more uniform as Lloyd's algorithm is iteratively applied. This process has been previously studied by \cite{klatt_universal_2019, hain_low-temperature_2020, hong_dynamical_2021}, in which quench-like behavior was observed: a rapid increase in order is followed by a dynamical arrest due to topological defects which freezes the system into a disordered hyperuniform state. Here, we further quantify this progression towards uniformity via both the probability distribution $p_k$ of the degree $k$ of the nodes (number of edges) and  the 
probability density function $f(\ell)$ of the 
lengths $\ell$ of the edges connecting nodes. In Fig.~\ref{fig:disorder}a, we observe that as $L$ increases the prevalence of $k=6$ nodes increases: the degree distribution $p_k$ becomes more strongly peaked around this value, as expected for an increasingly hexagonal lattice. In addition, a secondary peak gradually develops at $k=4$, driven by the existence of the square bounding box reducing the number of neighbors. Because there are more interior-nodes than boundary-nodes, the height of the $k=6$ peak is greater than that of the $k=4$ peak.  In Fig.~\ref{fig:disorder}b, we observe that as $L$ increases, there is a corresponding sharpening of the probability density function $f(\ell)$ around edge-length $\ell \approx 6$~mm, as edge lengths become more uniform across the network.  In Fig.~\ref{fig:disorder}c we show that, unlike the degree distribution, the edge distribution is invariant to the total number of points when length is rescaled by $\sqrt{N}$ to preserve the density of points within the bounding box.

We more directly quantify the disorder of the network through its information entropy $S$. From the histogram plotted in Fig.~\ref{fig:disorder}a, we calculate the degree entropy $S_k$ as the Shannon entropy
\begin{equation}\label{eq:degree_entropy}
S_k = -\sum_{k\in D} p_k \, \ln p_k
\end{equation}
where $D$ is the set of all possible degrees collected from the sample.
As we will see in Sec.~\ref{sec:Reff}, the effective resistance of the lattice is derived from a network description using edge weights proportional to the conductance $\sigma \propto 1/\ell$. Therefore, we calculate the edge entropy for a network weighted by $c \equiv 1/\ell$; because these values are drawn from a continuous distribution, we use the differential entropy 
\begin{equation}\label{eq:edge_entropy}
S_e = -\int_{\mathbb{R}} g(c) \, \ln g(c) \, dc
\end{equation}
where $g(c)$ is the probability density function of the edge-weights $c$, following the method outlined in \cite{vasicek_test_1976}.  This method utilizes the empirical cumulative distribution function, bypassing the need to choose a bin size for estimating the probability density function.  The code for performing these calculations is available on GitHub \cite{GitHub_config_stats}.

\begin{figure}
\includegraphics[width=0.9\linewidth]{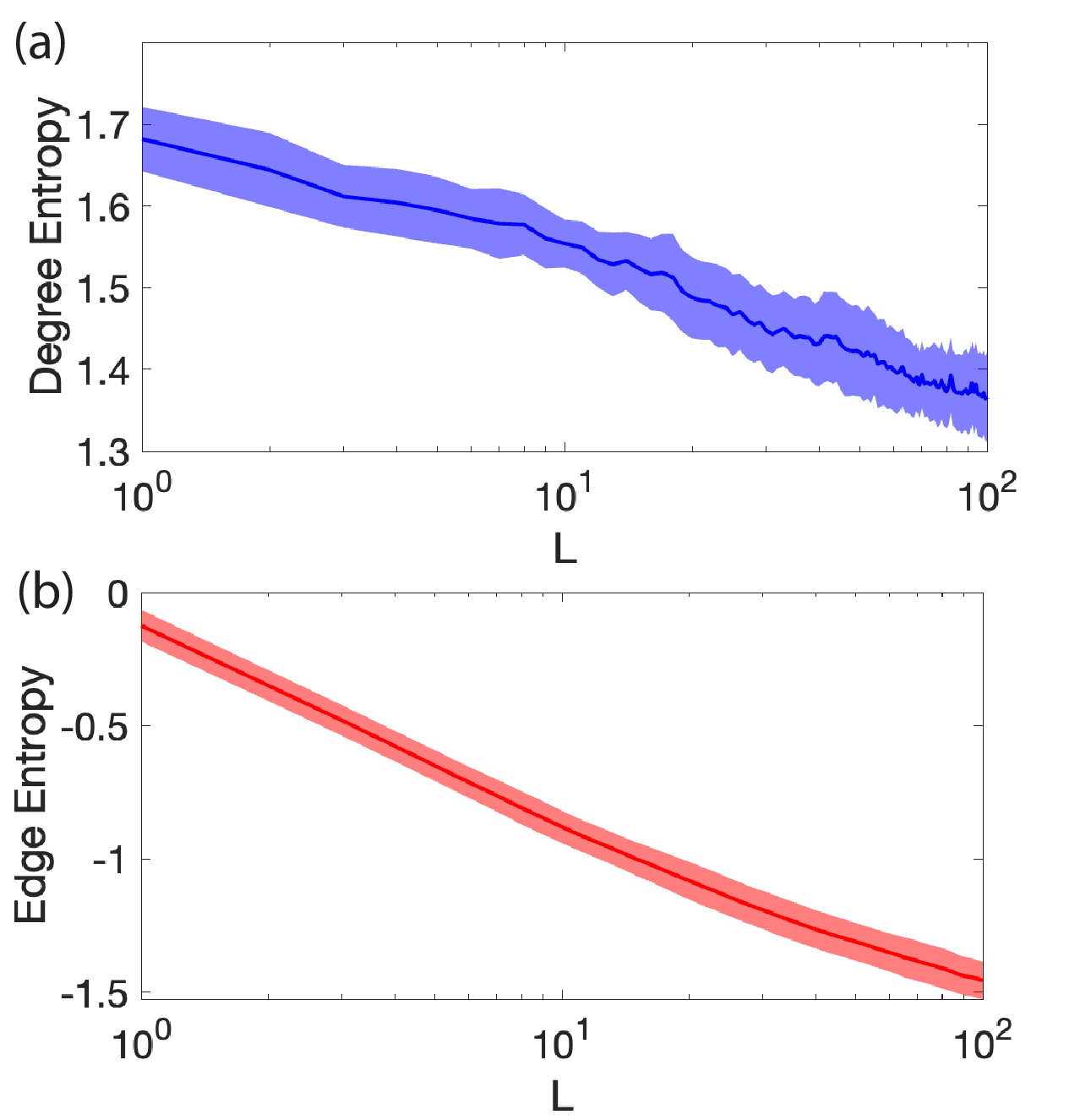}
\caption{Calculated for an ensemble of 20 realizations of an $N=200$ network, each subjected to Lloyd's iterations $L$:  (a) degree entropy $S_k$ calculated using Eq.~\eqref{eq:degree_entropy} and (b) edge entropy calculated using Eq.~\eqref{eq:edge_entropy}.  The solid lines are the averages over all 20 realizations, with the shaded region depicting $\pm$ one standard deviation. }
\label{fig:entropy}
\end{figure}

These two entropies are plotted in Fig.~\ref{fig:entropy} as a function of the number of iterations of Lloyd's algorithm. We observe that, as expected, the entropy decreases logarithmically as the configurations become more ordered at larger $L$. Both $S_k$ and $S_e$ have their most rapid decrease at low $L$, consistent with both the visual trend noted in Fig.~\ref{fig:4lloyds} and the changes in edge length distribution shown in Fig.~\ref{fig:disorder}b. Note that the differential entropy is not directly comparable with Shannon entropy; for example, the crystalline lattice network with a single value for the edge conductance would have zero Shannon entropy but infinitely negative differential entropy.

\subsection{Predicting effective resistance\label{sec:Reff}}

We compute the total effective resistance diagonally across the network configuration by assigning a resistance to each edge and then creating a system of linear equations using Kirchhoff's laws.  
This system of equations can be written in terms of a natural network descriptor, the combinatorial weighted graph Laplacian.
We summarize our derivation next to account for variable resistance along each edge, similar to \cite{ghosh_minimizing_2008}.  
It is an adaptation of the more common  $\mathcal{L}\vec{V}=r\vec{I}$ in the literature~(for example, see \cite{klein_resistance_1993,newman_finding_2004,newman_networks_2018}) where $\mathcal{L}$ is the combinatorial (unweighted) graph Laplacian, $r$ is a constant resistance for each edge, $\vec{I}$ represents the applied current to each node and $\vec{V}$ is the voltage at each node.

To model the physical samples, we assume each edge is a rectangular rod with a uniform cross-section and length equal to the distance between its endpoints.  
The resistance of the edge joining nodes $i$ and $j$ is therefore
\begin{equation}\label{eq:resistance}
    R_{ij} = \frac{\rho \ell_{ij}}{a}
\end{equation}
where $\ell_{ij}$ is the length of the edge, $\rho$ is the electrical resistivity of the material, and $a$ the cross-sectional area of the rod.
To arrive at a set of equations for the voltage at each node, we begin with
Kirchhoff's law that $V=IR$ across each edge, $I_{ij}=(V_i-V_j)/R_{ij}$, with $1/R_{ii}$ defined to be zero. Notice that $I_{ij} = - I_{ji}$ as expected.  Applying Kirchhoff's conservation of current at each node $i$, we have that
$\sum_{j=1}^N A_{ij}I_{ij} + I_i=0$, where $I_i$ represents the externally applied current to each node and $A_{ij}$ is the adjacency matrix of the network.  $A_{ij}=1$ if nodes $i$ and $j$ are connected and zero otherwise, thus the sum is over all nodes connected to node $i$.  Since 
 $I_{ij}=(V_i-V_j)/R_{ij}$, the above $N$ equations are equivalent to
\begin{equation}\label{eq:volt1}
\sum_{j=1}^N A_{ij} \frac{V_i - V_j}{R_{ij}} + I_i = 0 \quad \textrm{for }i=1\dots N.
\end{equation}
To model the experiment in orientation A(B), we find the NE(NW) clamped node and label it 1 and the SW(SE) clamped node and label it $N$.  We then set the externally applied current to be $I_1 = I_0, I_{N}=-I_0$ and $I_i=0$ otherwise.

The system in Eq.~\eqref{eq:volt1} can be put into matrix form by first 
defining a weighted adjacency matrix 
$
\tilde{A}_{ij} = \frac{1}{R_{ij}}
$
if nodes $i$ and $j$ are connected by an edge and 0 otherwise, and a weighted graph Laplacian
$\tilde{\mathcal{L}}_{ij} = (\delta_{ij}\sum_{j=1}^N \tilde{A}_{ij}) - \tilde{A}_{ij}$.  
Then the system in Eq.~\eqref{eq:volt1} is
$$
\tilde{\mathcal{L}} \vec V = \vec I
$$
The rank of a graph Laplacian is equal to the number of vertices $N$ minus the number of connected components in the graph, meaning for a connected graph like the ones studied here, the rank is always $N-1$.
Therefore, only the voltage difference is uniquely defined.
Taking $V_N=0$ without loss of generality, and solving the first $N-1$ equations we obtain the voltages at each node.  The effective resistance of the network is $\Reff = V_1/I_0$ and is independent of the choice of $I_0$.
Note that the current along each edge is obtained from the voltages using $I_{ij}=(V_i-V_j)/R_{ij}$. 
The code used for performing these calculations is available on GitHub  \cite{GitHub_config_resistance}.

Note that the resistivity $\rho$ and the cross-sectional area $a$ simply scale the graph Laplacian and therefore $\Reff$.  Forming the graph Laplacian matrix $\mathcal{L}_R$ using the edge-resistances in Eq.~\ref{eq:resistance} in which length is multiplied by $\rho/a$ is equivalent to forming the Laplacian matrix $\mathcal{L}_{\ell}$ from the edge-resistances given by their lengths $\ell_{ij}$ alone and then dividing by $\rho/a$:
\begin{equation}
\mathcal{L}_R = \frac{a}{\rho} \mathcal{L}_\ell.
\label{eq:rescale}
\end{equation}

\begin{figure}
\includegraphics[width=0.9\linewidth]{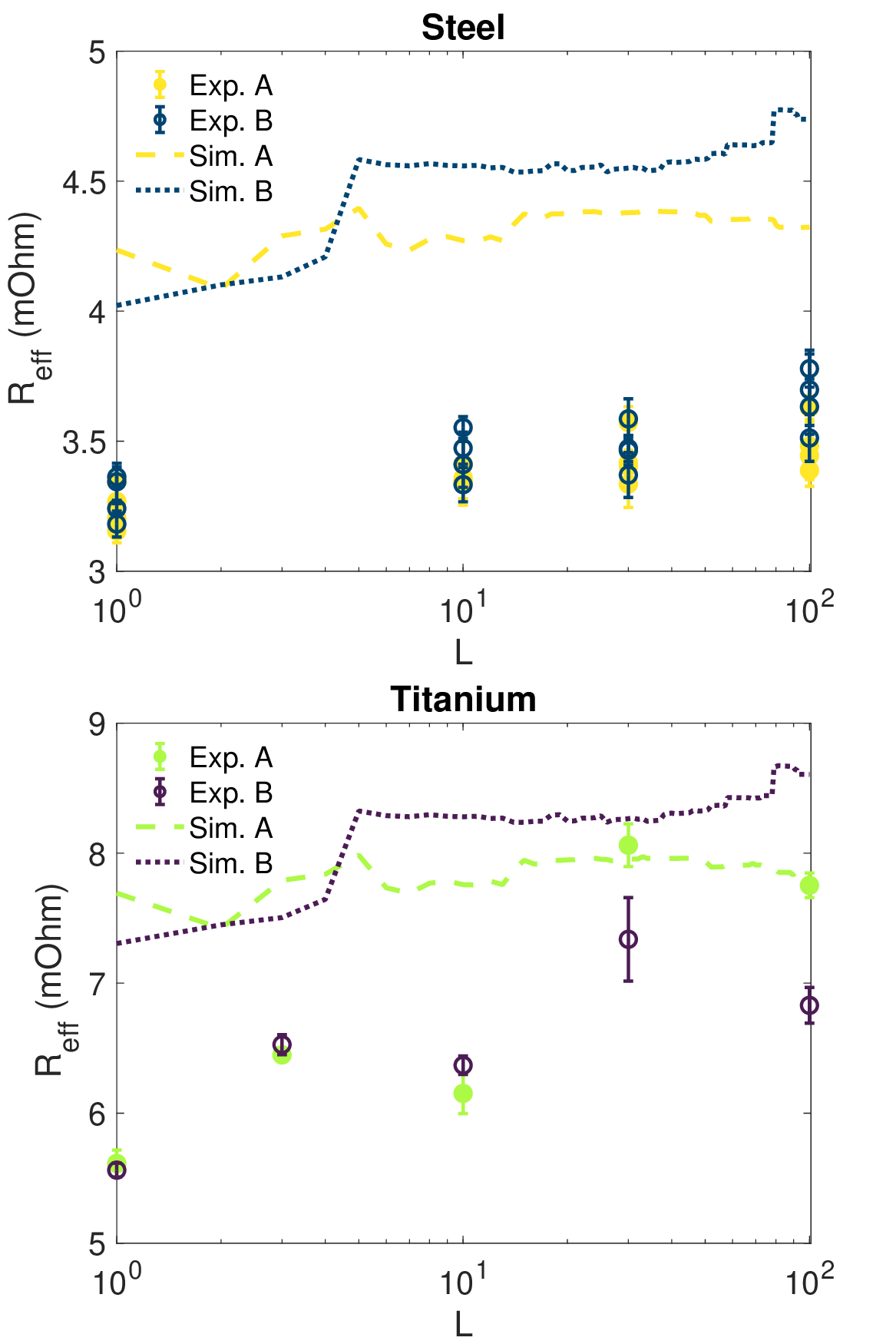}
\caption{$\Reff$ as a function of number of Lloyd's iterations, $L$, comparing both 4-point measurements (error bars) and predictions from the graph Laplacian (dashed line for orientation A and dotted line for orientation B) for both the steel 17-4PH and the titanium Ti-6Al-4V printed network configurations.
}
\label{fig:Reff}
\end{figure}

In Fig.~\ref{fig:Reff}, we compare these numerically computed $\Reff$ values to the values measured on physical samples, described in Sec.~\ref{sec:measure}. To compute the resistance of each edge according to Eq.~\eqref{eq:resistance}, we take the cross-sectional area $a = 0.03 \, \text{cm}^2$ for all samples. For 17-4PH steel the resistivity $\rho = 98 \, \mu\Omega\cdot$~cm  \cite{stashkov_magnetic_2019} and for Ti-6Al-4V titanium $\rho = 178  \, \mu\Omega\cdot$~cm \cite{nespoli_study_2022}. 
Using the single-bar samples of varying length described in Sec.~\ref{sec:measure}, we independently measure both $\rho$ and the contact resistance between the samples and the probes; we found values for $\rho$ consistent with the literature values (measured to be approximately 10\% higher) and with negligible contact resistance.  The lateral dimension of the bounding box (2000) corresponds to a printed length of $75$~mm; this ratio is used to convert the computational units for $\ell$ to cm. 

From the graph Laplacian method described above, we compute $\Reff$ for each iteration of Lloyd's algorithm applied to the 
point cloud realization used to create the printed samples. This result, rescaled to apply to steel vs. titanium using Eq.~\ref{eq:rescale}, and calculated for both orientations of the clamped nodes, is shown by the dashed (orientation A) and dotted (orientation B) lines in Fig.~\ref{fig:Reff}. With no free parameters, these lines reasonably reproduce the experimentally measured $\Reff$ values measured on the printed samples. The error bars on single data points show the standard deviation from 6 repeated measurements on a single sample, and the multiple data points at the same value of $L$ each come from different prints of the same network. Note that taking the measured values of $\rho$, rather than the literature values, does not resolve the $\sim 15\%$ observed here since the discrepancy is of the same magnitude, but in the opposite direction. A future study would need to resolve whether print-to-print variations of this magnitude are to be expected from the powder-bed process, or if there is a more fundamental cause involving transport on networks.

\subsection{Current anisotropy}

\begin{figure}
\includegraphics[width=\linewidth]{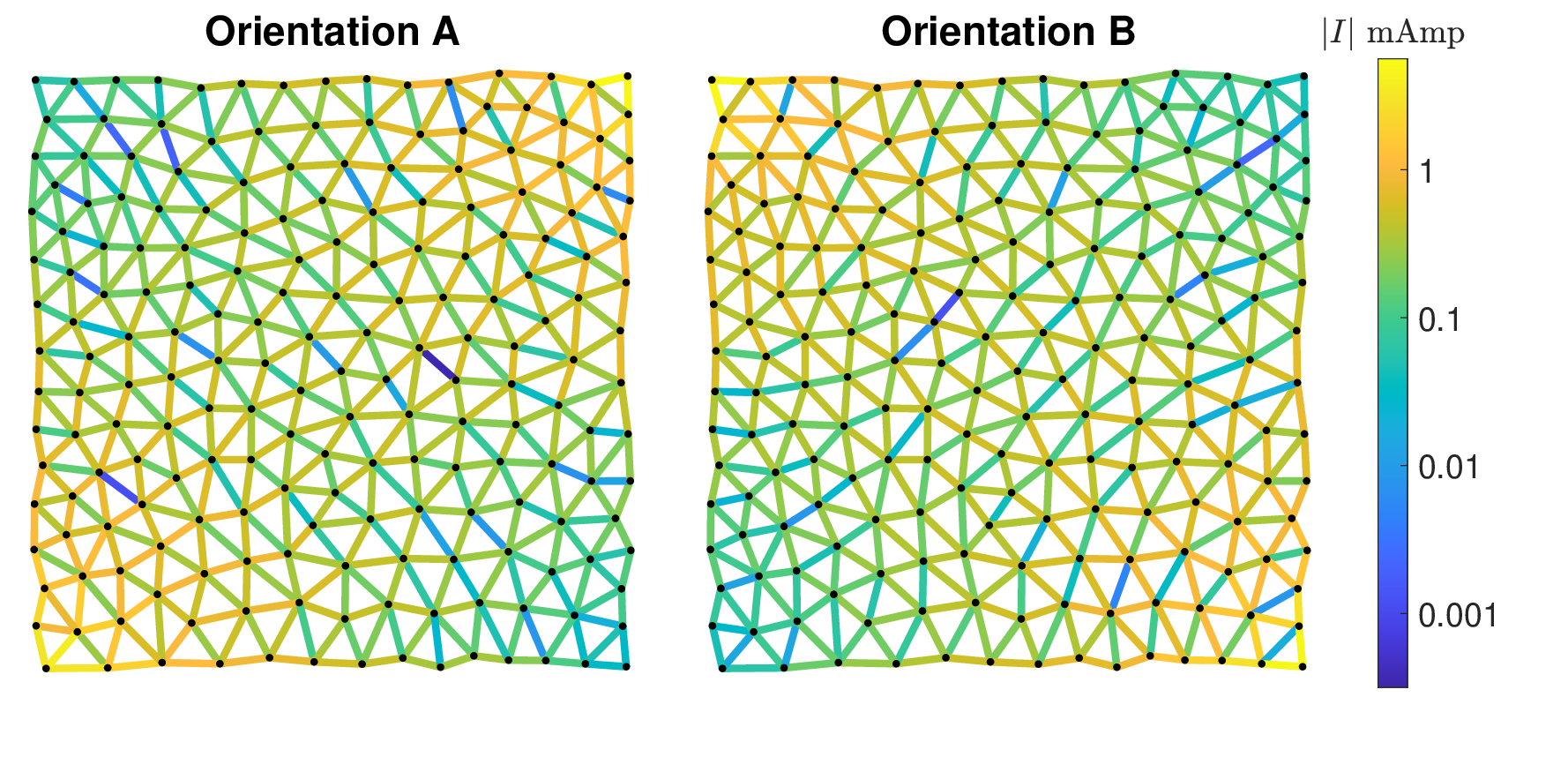}
\caption{The absolute value of the current flowing through each edge is shown on a logarithmic scale for the $L=10$ network from Fig.~\ref{fig:4lloyds}(b).
The current on specific edges depends on both the network topology and the direction of applied voltage. }
\label{fig:Current}
\end{figure}

From the numerical calculation of Sec.~\ref{sec:Reff}, we are able to obtain the current along each edge of the network. An example is shown in Fig.~\ref{fig:Current}, for both orientations of a single $L=10$ realization; the observations we make here hold for all realizations and $L$, whether ordered or disordered. Note that the absolute value of the current is depicted by the color bar since the positive direction is dictated by the order in which the nodes appear in the edge list representation of the matrix, rather than some physical meaning.  Colors are assigned to a logarithmic scale to provide visual contrast.  We observe that edges more aligned with the line connecting the clamped corners carry more current, as compared to those orthogonal to the direction of applied current. 
This anisotropy suggests that standard network characterizations such as centrality measures would not capture the global behavior of $\Reff$; indeed, we have yet to find a network measure that captures $\Reff$.

\subsection{Variability among the ensemble of realizations}

\begin{figure}
\includegraphics[width=0.9\linewidth]{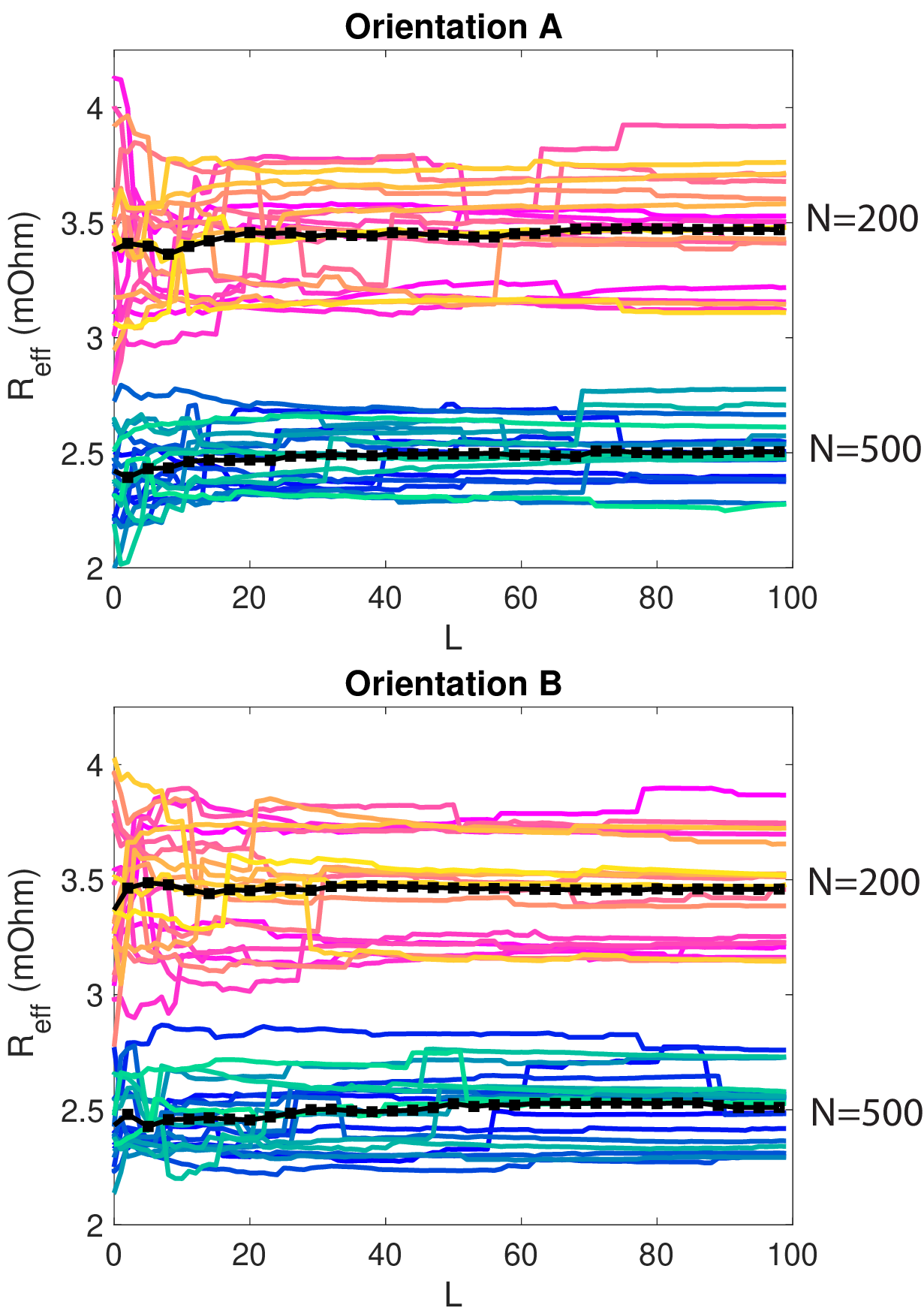}
\caption{Calculated $\Reff(L)$ for all 20 realizations, measured in both the A and B orientations of the clamped corners. Pink and yellow lines correspond to $N=200$ node networks; cyan and blue lines correspond to $N=500$ node networks.  Ensemble averages are shown with black lines and square dots.
}
\label{fig:Reff_ensemble}
\end{figure}

In Fig.~\ref{fig:Reff}, we observed that the numerically-computed value of $\Reff$ was in agreement with experimental measurements of the same configuration. However, while this particular realization showed $\Reff$ increasing with continued application of Lloyd's algorithm (progressing towards a more uniform configuration), this trend is inconsistent across the ensemble of initial point clouds taken as the seeds of the 20 independent realizations.
Fig.~\ref{fig:Reff_ensemble} shows the variability $\Reff(L)$ for the ensemble of realizations: each solid line follows one point cloud through iterations of Lloyd's algorithm. A mix of positive and negative slopes, both within a single progression of iterations and between samples, is the prevalent feature. Collectively, there is an observed decrease in the variability in $\Reff(L)$ for larger values of $L$. Since Fig.~\ref {fig:disorder}b and \ref{fig:entropy} both quantify a corresponding decrease in the disorder of the network, it appears that this is associated to a  decrease in variability of $\Reff$.

We also point out that the larger networks ($N=500$ nodes) have lower resistance than the smaller networks ($N=200$) nodes, due to resistors in parallel providing lower effective resistance. The ensemble of larger networks has less variability than the smaller network, at any given value of $L$.  A law of large numbers effect could explain this effect: within one network, there are more edges and therefore more paths between the two clamped corners, so it is less likely that the network has a more extreme path leading to larger or smaller resistance. As $L$ increases, 
both system sizes develop bands of common values of $\Reff$. This appears to be due to the partial crystallization of the network being frustrated by the bounding box of the domain.

\begin{figure}
    \includegraphics[width=\linewidth]{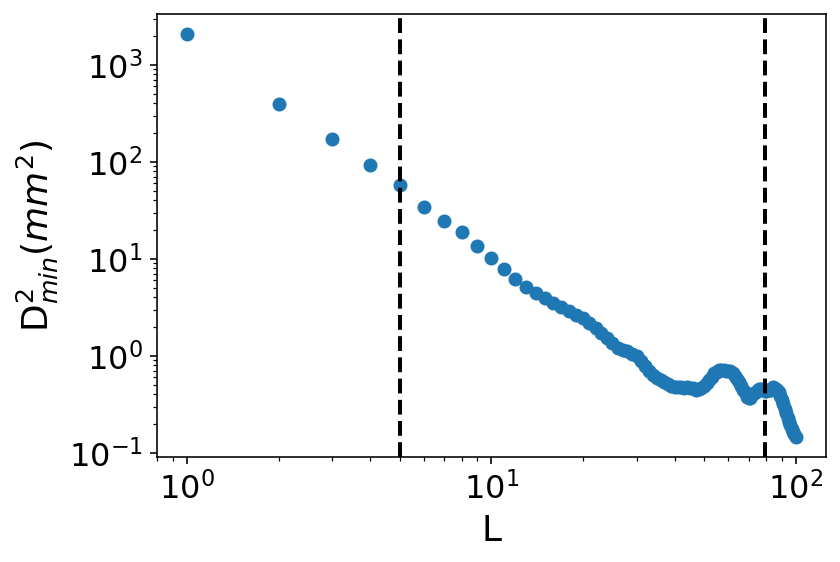}
\caption{The total $D^2_\textrm{min}$ for the network as a function of Lloyd's iteration~\cite{falk_dynamics_1998}, where the change in position is calculated between subsequent iterations. Vertical dashed lines at $L=5$ and $L=79$ mark the locations of the two jumps in $\Reff(L)$ shown in Fig.~\ref{fig:Reff}. A movie of the dynamics of the point cloud and of $D^2_\text{min}$ is available at \cite{youtube}.}
\label{fig:nonaffine}
\end{figure}

A common feature of the $\Reff(L)$ graphs is that we occasionally observe dramatic changes in $\Reff$ due to a single application of Lloyd's algorithm. In the example shown in Fig.~\ref{fig:Reff}, this occurs at $L=5$ and $L=79$, but only for the measurements done in Orientation B, and only for this particular point cloud. Other realizations exhibit similar jumps, as shown in Fig.~\ref{fig:Reff_ensemble}, and these commonly occur in only one of two clamping orientations. To seek an explanation, we quantified the rearrangements within the point cloud using the standard non-affinity measure $D^2_\text{min}$ \cite{falk_dynamics_1998}. As shown in Fig.~\ref{fig:nonaffine}, we observed that no significant rearrangements occurred at these two $L$ steps. Similarly, there are not any significant rearrangements of the Delaunay triangulation edges. This observation that no significant local change to the configuration is detected, together with the observed sensitivity to the orientation of the applied current, indicates that $\Reff$ is arising from the collective effects of the full network. Therefore, the lack of a network summary statistic capturing the changes in $\Reff$ is unsurprising.

\section{Discussion and conclusions}

We have developed a digital pipeline from algorithmically-created configurations with tunable disorder to printed materials; such materials open a large design space that is suitable for a variety of applications where tunability and low weight are important considerations. This methodology will additionally be applicable to  3D geometries, where disordered metamaterials research is less explored \cite{wit_simulation_2019,sniechowski_heterogeneous_2015} compared with designs based on repeating unit cells. While our original motivation had been to explore changes in bulk properties along an approach to hyperuniformity, we did not observe any strong effects. Instead, we observed an unexpected sensitivity to small configurational changes: jumps in $\Reff$ were present in samples regardless of the level of local disorder (progression of Lloyd's iterations towards a more uniform/crystalline state). 

We numerically analyzed our disordered networks as resistor networks to calculate the effective resistance via the graph Laplacian, finding that this method is able to quantitatively capture laboratory measurements of $\Reff$ on the printed samples. These direct comparisons demonstrate the feasibility of validating numerical results using a small number of printed samples, setting the stage for future optimization and design {\it in silico} rather than iterating through expensive prints.

Differences in the calculated and experimentally measured values may derive from process-related effects such as residual oxygen in solution, microstructural anisotropy, porosity, residual surface oxides, and roughness of the network beams. Oxygen in solution is a factor in additive metals which derive from powder feedstock and is known to have an influence on resistivity. The 17-4PH powder feedstock used in our study contained over 200
ppm as measured by inert gas fusion, compared with 52 ppm for a commercially obtained wrought bar.
While removing oxygen in bulk solids is challenging, an interesting future study would be 
 to resolve surface effects on each connecting beam by performing modifications such as chemical or micro-abrasion to remove beam oxide layers and reduce surface roughness of the printed networks.

The freeform nature of additive manufacturing unlocks new algorithm-based design methodologies compared with traditional CAD modeling.  This pipeline adds to the toolbox of recently established innovative digital designs such as triply periodic minimized surfaces \cite{dutkowski_review_2022,sharma_additively_2022,al-ketan_multifunctional_2019, praveen_kumar_open-source_2020}, unit cell lattice structures \cite{van_hooreweder_advanced_2017,askari_additive_2020}, Voronoi tesselations \cite{bahr_novel_2017, herath_mechanical_2021,hooshmand-ahoor_mechanically-grown_2022, martinez_polyhedral_2018}, as well as generative modeling and porous materials \cite{hsu_generative_2022,ullah_system_2020} to produce non-traditional geometries for applications ranging from heat exchangers to medicine. 

For both the single printed sample, and the ensemble of samples studied {\em in silico} we
observed that common network measures successfully characterize the degree of disorder as a function of the number of iterations of Lloyd's algorithm, as the point cloud configurations progress from disordered to ordered. 
Surprisingly, we saw no strong dependence of $\Reff$ on the degree of disorder other than a decrease in variability across the ensemble as the degree of disorder decreased.
This observation appears at odds with both 
theoretical \cite{klatt_wave_2022} and laboratory \cite{aubry_experimental_2020} studies, that photonic band gaps are highly sensitive to the degree and type of hyperuniformity.
One possible explanation is that measures of disorder, like the width of the edge distribution or entropy employed here, do not necessarily detect hyperuniformity.  Another possible explanation is that not all types of transport are sensitive to the hyperuniformity of a material, which should be explored in future studies.

Our inability to find 
a summary network statistic that could replace the full graph Laplacian (exact solution) suggests that effective resistance is more sensitive to the full network topology than one might think.
One challenge to finding such a metric is that we observed a strong anisotropy due to the imposed direction of the driving current, yet topological network measures contain no notion of directionality. 
The network measure that showed the most promise was the shortest path length, which does have directionality as it measures network length between two specific nodes (other measures are averaged over the entire network).
However, we did not include it in this paper as its effect is subtle, with no clear trend on average.  Future studies will further investigate this statistic and directional versions of other statistics created by changing the way they are averaged over the network.

Furthermore, jumps in $\Reff$ occurred at different steps along the evolution of the configuration, depending on which orientation was chosen for applying the current, and these jumps  were not accompanied by large changes in the point cloud configuration as measured by $D_{\min}^2$.  We noticed some correlation between changes in the adjacency matrix (adding or removing edges) and jumps in $\Reff$, but not all changes in the network structure were accompanied by equally-sized jumps in $\Reff$.  While the Bauer-Fike Theorem bounds changes in the eigenvalues of a matrix by changes to the matrix itself, there is no clear link between the eigenvalues of the graph Laplacian and the measured $\Reff$.  
A promising approach may be to instead calculate the total effective resistance, which is the average effective resistance taken over all pairs of nodes in the network.  This is known to be proportional to the sum of the inverse of the eigenvalues of the graph Laplacian, but is experimentally intractable.
We will continue to investigate these intriguing phenomena exposed in this work, seeking to better understand directionally-dependent network characterizations like $\Reff$ and shortest path length versus direction-insensitive characterizations that are averaged over the entire network.

This work established the feasibility of manufacturing digital designs, creating a pathway to further research on disordered metamaterials, and the development of new characterization metrics.

\section*{Acknowledgements}

This work is supported by the collaborative NSF DMREF grant numbers CMMI-2323341 and CMMI-2323342 and NSF grant number DMS-2307297.  The authors thank Charles Maher for useful discussions on hyperuniformity.

%

\end{document}